ARTICLE

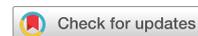

# A simple method for measuring inequality

Thitithep Sitthiyot[1✉] & Kanyarat Holasut[2]

**ABSTRACT**   To simultaneously overcome the limitation of the Gini index in that it is less sensitive to inequality at the tails of income distribution and the limitation of the inter-decile ratios that ignore inequality in the middle of income distribution, an inequality index is introduced. It comprises three indicators, namely, the Gini index, the income share held by the top 10%, and the income share held by the bottom 10%. The data from the World Bank database and the Organization for Economic Co-operation and Development Income Distribution Database between 2005 and 2015 are used to demonstrate how the inequality index works. The results show that it can distinguish income inequality among countries that share the same Gini index but have different income gaps between the top 10% and the bottom 10%. It could also distinguish income inequality among countries that have the same ratio of income share held by the top 10% to income share held by the bottom 10% but differ in the values of the Gini index. In addition, the inequality index could capture the dynamics where the Gini index of a country is stable over time but the ratio of income share of the top 10% to income share of the bottom 10% is increasing. Furthermore, the inequality index could be applied to other scientific disciplines as a measure of statistical heterogeneity and for size distributions of any non-negative quantities.

[1] Chulalongkorn University, Bangkok, Thailand. [2] Khon Kaen University, Khon Kaen, Thailand. ✉email: thitithep@cbs.chula.ac.th

## Introduction

The Gini index was devised by an Italian statistician named Corrado Gini in 1912. By far, it has arguably been the most popular measure of socioeconomic inequality, especially in income and wealth distribution, given that there are well over 50 inequality indices as reported in Coulter (1989) (see Eliazar, 2018; McGregor et al., 2019 for recent updates on the inequality measures). The use of the Gini index is not limited to the field of socioeconomics, however. According to Eliazar and Sokolov (2012), the application of the Gini index has grown beyond socioeconomics and reached various disciplines of science. Examples include astrophysics—the analysis of galaxy morphology (Abraham et al., 2003); ecology—patterns of inequality between species abundances in nature and wealth in society (Scheffer et al., 2017); econophysics—wealth inequality in minority game (Ho et al., 2004); scale invariance in the distribution of executive compensation (Sitthiyot et al., 2020); engineering—the analysis of load feature in heating, ventilation, and air conditioning systems (Zhou et al., 2015); finance—the analysis of fluctuations in time intervals of financial data (Sazuka and Inoue, 2007); human geography—measuring differential accessibility to facilities between various segments of population (Cromley, 2019); informetrics—the analysis of citation (Bertoli-Barsotti and Lando, 2019); medical chemistry—the analysis of kinase inhibitors (Graczyk, 2007); population biology—heterogeneities in transmission of infectious agents (Woolhouse et al., 1997); public health—the analysis of life expectancy (De Vogli et al., 2005); the analysis of real biological harm (Sapolsky, 2018); renewable and sustainable energy—the analysis of irregularity of photovoltaic power output (Das, 2014); sustainability science—the study of land change (Rindfuss et al., 2004); transport geography—equity in accessing public transport (Delbosc and Currie, 2011); selection of tram links for priority treatments (Pavkova et al., 2016). In effect, the Gini index is applicable to any size distributions in the context of general data sets with non-negative quantities such as count, length, area, volume, mass, energy, and duration (Eliazar, 2018). However, in order to demonstrate our method for measuring inequality, we focus our analysis on the subject of income.

The Gini index can be derived from the Lorenz curve framework (Lorenz, 1905), which plots the Cartesian coordinates where the abscissa is the cumulative normalized rank of income from the lowest to the highest ($x$) and the ordinate is the cumulative normalized income from the lowest to the highest ($y$) as illustrated in Fig. 1. According to Gini (2005), the Gini index can be calculated as the ratio of the area between the perfect equality line and the Lorenz curve ($A$) divided by the total area under the perfect equality line ($A + B$). The Gini index takes values in the unit interval. The closer the index is to zero (where the area $A$ is small), the more equal the distribution of income. The closer the index is to one (where the area $A$ is large), the more unequal the distribution of income.

The advantage of the Gini index is that inequality of the entire income distribution can be summarized by using a single statistic that is relatively easy to interpret since it takes values between 0 and 1. This allows for comparison among countries with different population sizes. In addition, the data on the Gini index is easy to access, regularly updated and reported by countries and international organizations. Despite its advantages as a statistical measure of income inequality, Atkinson and Bourguignon (2015) note that a country with lower Gini index does not always imply that income distribution in that country is more equal than that of a country with higher Gini index. This is because the Lorenz curves of the two countries may intersect, reflecting different income distributions.

To obtain a complete ranking of and to quantify the difference in income inequality among countries, Atkinson (1970) devises a social welfare-based inequality index as follows:

$$A(\varepsilon) = 1 - \left(\frac{1}{N}\sum_{i=1}^{N}\left(\frac{y_i}{\bar{y}}\right)^{1-\varepsilon}\right)^{\frac{1}{1-\varepsilon}}, \varepsilon \neq 1 \quad (1)$$

$$A(\varepsilon) = 1 - \frac{\prod_{i=1}^{N}\left(y_i^{\left(\frac{1}{N}\right)}\right)}{\bar{y}}, \quad \varepsilon = 1 \quad (2)$$

where $y_i$ denotes the individual income, $\bar{y}$ denotes the average income, $N$ is the number of population, and $\varepsilon$ is the inequality aversion parameter. This index takes values between 0 and 1. The cornerstone of the Atkinson index is the concept of equally distributed equivalent level of income ($y_{EDE}$), which is defined as the percentage of total income that a given society would have to forego in order to have more equal shares of income among individuals in that society. Note that if $\varepsilon \neq 1$, $y_{EDE} = \left(\frac{1}{N}\sum_{i=1}^{N}y_i^{1-\varepsilon}\right)^{\frac{1}{1-\varepsilon}}$. When $\varepsilon = 1$, $y_{EDE} = \prod_{i=1}^{N}(y_i)^{\frac{1}{N}}$.

The notion of $y_{EDE}$ depends on the degree of inequality aversion parameter $\varepsilon$, which technically could range between 0 and $\infty$. As $\varepsilon$ increases, a society attaches more weight to transfers at the lower end of the income distribution and less weight to transfers at the top. By using $\varepsilon = 2$, Atkinson (1970) finds that, of the 50 pairwise comparisons where the Lorenz curves intersect, his inequality index would disagree with the Gini index in seventeen cases. For a lower degree of inequality aversion with $\varepsilon = 1$, there are only five cases that would disagree with the Gini index. According to Atkinson (1970), the Gini index tends to give the rankings that are similar to those reached with a relatively low degree of inequality aversion.

Although the advantages of the Atkinson index are that it provides a complete ranking of income distributions and makes explicit the social welfare function underlying the income inequality measure, which could be useful for policy decisions, Cowell (2011) and McGregor et al. (2019) note that the ranking of income distributions can vary widely depending upon the choice of social welfare functions and the intrinsic aversion to inequality, which may not necessarily be the same among countries.

To avoid the social welfare judgment, a class of generalized entropy (GE) indices can be used as an alternative measure for ranking income inequality when the Lorenz curves of the two

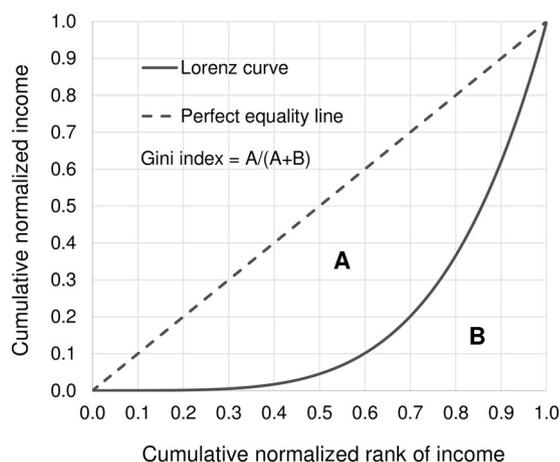

**Fig. 1 Lorenz curve.** The Gini index is calculated as the ratio of the area between the perfect equality line and the Lorenz curve ($A$) divided by the total area under the perfect equality line ($A + B$).

countries intersect. The GE index is defined as follows:

$$\text{GE}(\alpha) = \frac{1}{\alpha(\alpha-1)} \left[ \frac{1}{N} \sum_{i=1}^{N} \left(\frac{y_i}{\bar{y}}\right)^{\alpha} - 1 \right] \quad (3)$$

The theoretical values of the GE($\alpha$) index vary between 0 and $\infty$, with 0 representing equal income distribution and higher values representing higher levels of income inequality. The GE($\alpha$) index as shown in Eq. (3) defines a class because it assumes different forms depending upon the value assigned to the parameter $\alpha$, which is a weight given to inequalities in different parts of the income distribution. The less positive the parameter $\alpha$ is, the more sensitive the index is to inequalities at the bottom of the income distribution while the more positive the parameter $\alpha$ is, the more sensitive the index is to inequalities at the top (Bellù and Liberati, 2006). Bellù and Liberati (2006) also note that, in principle, the parameter $\alpha$ can take any real values from $-\infty$ to $\infty$. However, from a practical point of view, $\alpha$ is normally chosen to be positive. This is because, for $\alpha < 0$, this class of indices is undefined if there are zero incomes. GE(0) is referred to as the mean logarithmic deviation, which is defined as follows:

$$\text{GE}(0) = \frac{1}{N} \sum_{i=1}^{N} \ln\left(\frac{\bar{y}}{y_i}\right) \quad (4)$$

GE(1) is known as the Theil inequality index, named after the author who devised it in 1967. The Theil index is defined as follows:

$$\text{GE}(1) = \frac{1}{N} \sum_{i=1}^{N} \frac{y_i}{\bar{y}} \ln\left(\frac{y_i}{\bar{y}}\right) \quad (5)$$

While a class of GE($\alpha$) indices can overcome the limitation of the Gini index in ranking income inequality when the Lorenz curves of the two countries cross, it should be noted that the exact specification of the GE($\alpha$) index depends upon the value of $\alpha$, which may vary from country to country, making it difficult to compare income inequality among different countries.

Another limitation of using the Gini index is whenever two or more countries share the same value of the Gini index but income inequality among them could be very different if taking into consideration the information on the income share held by the richest and that held by the poorest. For example, based on the data from the World Bank, in 2015, Greece and Thailand have the same Gini index (0.360) but the ratio of the income share held by the richest 10% to the income share held by the poorest 10% in Greece is 13.8 while that in Thailand is 8.9. In addition, according to the Organization for Economic Co-operation and Development Income Distribution Database (OECD IDD), it shows that, in 2015, the United Kingdom and Israel also share the same Gini index (0.360) but the ratio of the income share of the top 10% to the income share of the bottom 10% in the United Kingdom equals 4.2, whereas that in Israel equals 5.8. That countries share the same Gini index but differ in the income gap between the richest and the poorest indicates that the Gini index alone cannot tell the difference in income inequality among countries.

Furthermore, Atkinson (1970) notes that the Gini index is more sensitive to changes in the middle of income distribution and less sensitive to changes at the top and the bottom of income distribution. Palma (2011) analyses income inequality across countries using the inter-decile ratios and finds that the rising in income inequality comes from an increased diversity in the income share held by the top 10% and the income share held by the bottom 40% while the income share of the deciles 5 to 9 remains stable over time. According to Palma (2011), countries with high income inequality are simply those in which the top 10% are more successful at subsidizing their insatiable appetite with the income of the bottom 40%. Palma (2011) suggests that, in order to reduce income inequality, policymakers should direct policies towards lowering the ratio of the income share held by the top 10% over the income share held by the bottom 40%.

Besides the ratio of the income share of the top 10% over the income share of the bottom 40% as proposed by Palma (2011), there are other inter-decile ratios that emphasize the tails of income distribution, for example, the ratio of the income share held by the top 10% to the income share held by the bottom 10% and the ratio of the income share held by the top 20% to the income share held by the bottom 20%. The data used to calculate these inter-decile ratios or the ratios themselves including the Palma index are regularly updated and reported along with the Gini index by international organizations, such as the World Bank, the OECD, and the Human Development Report Office as the measures of income inequality.

While these inter-decile ratios seem to be easy to understand and convey the information to public and policymakers with regard to income inequality, it should be noted that if carefully considering the values that these ratios can take, we would find that they are between ¼ and $\infty$ for the Palma index, and between 1 and $\infty$ for both the ratio of the income share of the top 10% to the income share of the bottom 10% and the ratio of the income share of the top 20% to the income share of the bottom 20%. From a mathematical and practical point of view, these values are more difficult to interpret and compare among countries since they have no upper bound relative to other inequality indices whose values are bounded. The same argument could be applied for a class of GE($\alpha$) indices as discussed earlier. As noted in Eliazar (2018), indices whose values are bounded are much more tangible to human perception than those whose values are unbounded. In addition, by construction, these inter-decile ratios capture income inequality between the top and the bottom of distribution and ignore income of those in the middle of distribution.

To overcome the limitations of the Gini index and the inter-decile ratios as discussed above, we devise an alternative method for measuring inequality. Our method is quite simple. It utilizes the Gini index, the income share held by the top 10%, and the income share held by the bottom 10% to construct a composite index. These three indicators comprising the inequality index are selected based on availability, accessibility, and continuity of the data without the need to collect the data on income distribution at the micro level. We are well aware that the accuracy of the data on the Gini index and the income shares depends on the population survey methods and/or the probability laws governing the distribution of income as discussed in Eliazar and Sokolov (2012), Chakrabarti et al. (2013), and Sarabia et al. (2019). While research on these issues is continuing, we hope that our simple method for measuring inequality would be useful not only for socioeconomics but also for other disciplines of science as a measure of statistical heterogeneity and for general size distributions, providing ones have the data on the Gini index and the income share of the top 10% and that of the bottom 10% or the respective share's ratio of any non-negative quantities.

## Methods

To derive our inequality index for any given country i ($I_i$), let $\text{Gini}_i$ denote the Gini index of country i (in decimal places) and $\left(\frac{B_{10}}{T_{10}}\right)_i$ be the ratio of the income share held by the bottom 10% ($B_{10}$) to the income share held by the top 10% ($T_{10}$) in country i. In addition, let $H_i = 1 - \left(\frac{B_{10}}{T_{10}}\right)_i^{\alpha}$, $0 \leq H_i \leq 1$. The exponent alpha ($\alpha$) is a weight such that $\text{avg. Gini} = 1 - \left(\text{avg.}\left(\frac{B_{10}}{T_{10}}\right)\right)^{\alpha}$, where avg. Gini is the average value of all countries' Gini index in the sample whereas avg.$(\frac{B_{10}}{T_{10}})$ is the average value of all countries' $\frac{B_{10}}{T_{10}}$ ratio from the same sample. Note that the justification for the parameter $\alpha$ is to rescale the weight of the $H_i$ so that it is more or less in balance with that of the Gini$_i$. The higher the value of $\alpha$ is, the closer the value of the $H_i$ is to 1, meaning that more weight is assigned to the $H_i$ and less weight is assigned to the Gini$_i$. In contrast, the lower the value of $\alpha$ is, the closer the value of the $H_i$ is to 0, implying that we assign more weight on the Gini$_i$ and less weight on the $H_i$.

Using the empirical data on the Gini index and on the $\frac{B_{10}}{T_{10}}$ ratio, the parameter $\alpha$ can be calculated as $\frac{\ln(1-\text{avg. Gini})}{\ln(\text{avg.}(\frac{B_{10}}{T_{10}}))}$. In practice, the value of $\alpha$ used to calculate the $H_i$ would vary from sample to sample. This would make the value of $\alpha$ to be sample dependent, resulting in the values of $H_i$ for countries in the sample to be incomparable across periods. The alternative is to calculate the average value of $\alpha$ ($\bar{\alpha}$) across all samples, set it as a constant, and then use it for the calculation of all $H_i$'s regardless of the countries and the years being investigated. In this way, the parameter $\alpha$ would be standardized and sample independent since, from now on, we only need the Gini index and the ratio of income shares of countries of interest in any given years to calculate our inequality index. Finally, by applying the Pythagoras' theorem, the inequality index for any given country i ($I_i$) can be written as a function of the Gini$_i$ and the $H_i$ as follows[1]:

$$I_i = f(\text{Gini}_i, H_i) = \frac{\sqrt{\text{Gini}_i^2 + H_i^2}}{\sqrt{2}}, \quad 0 \leq I_i \leq 1 \qquad (6)$$

Our inequality index ($I$) takes values in the unit interval where the closer the index is to zero, the more equal the distribution of income and the closer the index is to one, the more unequal the distribution of income.

To demonstrate our method, we use the annual data on the Gini index and the income shares in 2015 from the World Bank (2019a, 2019b, 2019c) containing 75 countries and from the OECD IDD (2019a, 2019b) comprising 35 countries. The reason to use the data in 2015 is that it has more countries than those in 2016, 2017, and 2018. This would allow us to have more chance to find countries that have the same value of the Gini index but differ in the $\frac{T_{10}}{B_{10}}$ ratios, and countries that share the same $\frac{T_{10}}{B_{10}}$ ratio but differ in values of the Gini index, all of which would be used as examples to verify our method. In addition, the data from the World Bank and the OECD IDD between 2005 and 2015 are employed in order to calculate $\bar{\alpha}$ for the entire period, and, more importantly, to show that our inequality index could capture the case where countries whose the Gini index is stabilizing or falling across time but the ratio of the income share of the top 10% to that of the bottom 10% is increasing.

## Results

We first calculate the descriptive statistics of the $\frac{B_{10}}{T_{10}}$ ratio, the Gini index, and the $H$ as well as the correlation coefficients between these three indicators using the data from the World Bank and the OECD IDD from 2005 to 2015 (see Tables S1–S44 in the Supplementary Information). The results shown in the correlation matrix indicate that the Gini index is positively correlated with the $H$ while the $\frac{B_{10}}{T_{10}}$ ratio is negatively correlated with the Gini index and the $H$. In all cases, the correlation coefficients, in absolute values, are > 0.900.

Next, we calculate the values of the weight $\alpha$ in each year for the entire 11-year sample from both databases. We find that, for the World Bank database, the value of $\alpha$ is between 0.197 and 0.207. For the OECD IDD, the value of $\alpha$ ranges between 0.271 and 0.281. We then calculate the value of $\bar{\alpha}$ across all samples from both databases and find that $\bar{\alpha} = 0.239$. The values of parameter $\alpha$ and $\bar{\alpha}$ calculated using the World Bank database and the OECD IDD between 2005 and 2015 are reported in Tables S45 and S46 in the Supplementary Information.

Although we could fix the value of $\bar{\alpha} = 0.239$ as a constant and use it to calculate the $H_i$ as discussed in the "Methods", for the purpose of standardizing and using it in practice, this number is not very easy to work with practically and mathematically; imagine memorizing the number and taking a root of 0.239. After a careful consideration, without losing its function as a parameter that balances the weight of the $H_i$ with that of the Gini$_i$, we would like to propose the value of $\alpha$ to be 0.25 or ¼ and set it as a constant for the ease of the calculation of the $H_i$, where $H_i = 1 - (\frac{B_{10}}{T_{10}})_i^{\frac{1}{4}}$, irrespective of the countries and/or the years being studied. From our viewpoint, the $\alpha$ value of 0.25 or ¼ is much easier to use practically and mathematically; imagine taking a square root twice, which could be done by using a simple calculator compared to taking a root of 0.239. In addition, the value of $\alpha$ that we propose is not very far off from the empirical value of $\bar{\alpha}$ calculated from the two databases (0.25 vs. 0.239). For these reasons, we would like to define our inequality index ($I_i$) as follows:

$$I_i = \frac{\sqrt{\text{Gini}_i^2 + \left[\left(1 - (\frac{B_{10}}{T_{10}})_i\right)^{\frac{1}{4}}\right]^2}}{\sqrt{2}}, \quad 0 \leq I_i \leq 1 \qquad (7)$$

Given the data on the Gini index (Gini$_i$) and the calculated $H$ value for each country ($H_i$) using $\alpha = 0.25$ or ¼ as a constant, we can compute the inequality index ($I_i$) for that country.

Table 1 presents the results of our ranking of income inequality based on the Gini index using the World Bank database in 2015. The results indicate that the inequality index ($I$) can differentiate income inequality in case two or more countries share the same Gini index but differ in the income gap between the top and the bottom. As discussed in the "Introduction", Greece and Thailand share the same level of income inequality if measured by the Gini index (0.360). However, the $\frac{T_{10}}{B_{10}}$ ratio in Greece = 13.8 while that in Thailand = 8.9, indicating that Greece has higher income inequality than Thailand, which cannot be explained by the Gini index. However, our inequality index ($I$) can tell this difference since Thailand has $I = 0.391$, whereas Greece has $I = 0.425$. Thus, using our inequality index ($I$), we can say that Greece has a higher level of income inequality than Thailand. The evolution of the Gini index, the $\frac{T_{10}}{B_{10}}$ ratio, and the inequality index ($I$) of Greece and Thailand is shown in Fig. 2. Our results in Table 1 also show that when comparing the rankings of income inequality among countries using our inequality index ($I$) with those using the Gini index, there are 62 countries that their rankings have been changed while there are 13 countries whose rankings remain the same.

In addition, our inequality index ($I$) would be able to distinguish income inequality of two or more countries that have the same $\frac{T_{10}}{B_{10}}$ ratio but have different values of the Gini index. Using the World Bank database, in 2015, Malta and Slovak Republic share the same $\frac{T_{10}}{B_{10}}$ ratio (6.74) but Malta has the Gini index = 0.294 while the Gini index of Slovak Republic = 0.265, indicating that income inequality in Malta is higher than that in Slovak Republic as measured by the Gini index. Our inequality index ($I$) suggests the same results since the inequality index ($I$) of Malta = 0.339, whereas that of Slovak Republic = 0.327. The evolution of the Gini index, the $\frac{T_{10}}{B_{10}}$ ratio, and the inequality index ($I$) of Malta and Slovak Republic is illustrated in Fig. 3.

Furthermore, our inequality index ($I$) would be able to capture the case where the Gini index of a country is stabilizing across time but the income gap between the top 10% and the bottom 10% is rising. According to the World Bank database, during 2008 and 2014, the Gini index of Mexico is around 0.453 but the ratio of $\frac{T_{10}}{B_{10}}$ is rising from 17.6 in 2008 to 18.6 in 2014. This suggests that the income inequality in Mexico is increasing if measured by the $\frac{T_{10}}{B_{10}}$ ratio but is not if measured by the Gini index. By combining the Gini index and the $\frac{T_{10}}{B_{10}}$ ratio in order to construct a single composite measure, our inequality index (I) could capture this dynamics since the value of our inequality index (I) shows a rising trend from 0.480 in 2008 to 0.489 in 2014. Figure 4 illustrates the evolution of the Gini index, the $\frac{T_{10}}{B_{10}}$ ratio, and the inequality index (I) of Mexico.

Table 1 Income inequality indicators for 75 countries in 2015 (the World Bank database).

| Ranking by Gini index | Country | Gini index | Inequality index (I) | $T_{10}/B_{10}$ | $H$ |
| --- | --- | --- | --- | --- | --- |
| 1 | Slovenia | 0.254 | 0.302 | 5.38 | 0.344 |
| 2 | Ukraine | 0.255 | 0.298 | 5.14 | 0.336 |
| 3 | Belarus | 0.256 | 0.299 | 5.17 | 0.337 |
| 4 | Czech Republic | 0.259 | 0.309 | 5.67 | 0.352 |
| 5 | Slovak Republic | 0.265 | 0.327 | 6.74 | 0.379 |
| 5 | Kosovo | 0.265 | 0.307 | 5.38 | 0.343 |
| 7 | Kazakhstan | 0.268 | 0.306 | 5.26 | 0.340 |
| 8 | Moldova | 0.270 | 0.311 | 5.49 | 0.347 |
| 9 | Finland | 0.271 | 0.315 | 5.74 | 0.354 |
| 10 | Norway | 0.275 | 0.326 | 6.37 | 0.371 |
| 11 | Belgium | 0.277 | 0.329 | 6.53 | 0.374 |
| 12 | Denmark | 0.282 | 0.330 | 6.43 | 0.372 |
| 12 | Netherlands | 0.282 | 0.332 | 6.57 | 0.375 |
| 14 | Serbia | 0.285 | 0.329 | 6.24 | 0.367 |
| 15 | Kyrgyz Republic | 0.290 | 0.328 | 6.05 | 0.362 |
| 16 | Sweden | 0.292 | 0.349 | 7.63 | 0.398 |
| 17 | Malta | 0.294 | 0.339 | 6.74 | 0.379 |
| 18 | Hungary | 0.304 | 0.358 | 7.93 | 0.404 |
| 19 | Austria | 0.305 | 0.358 | 7.93 | 0.404 |
| 20 | Croatia | 0.311 | 0.367 | 8.59 | 0.416 |
| 21 | Germany | 0.317 | 0.364 | 8.00 | 0.405 |
| 22 | Egypt, Arab Rep. | 0.318 | 0.355 | 7.13 | 0.388 |
| 22 | Ireland | 0.318 | 0.366 | 8.19 | 0.409 |
| 22 | Poland | 0.318 | 0.360 | 7.61 | 0.398 |
| 25 | Switzerland | 0.323 | 0.365 | 7.88 | 0.403 |
| 26 | Armenia | 0.324 | 0.365 | 7.85 | 0.403 |
| 27 | Estonia | 0.327 | 0.378 | 9.04 | 0.423 |
| 27 | France | 0.327 | 0.374 | 8.58 | 0.416 |
| 29 | Tunisia | 0.328 | 0.369 | 8.00 | 0.405 |
| 30 | United Kingdom | 0.332 | 0.378 | 8.76 | 0.419 |
| 31 | Pakistan | 0.335 | 0.366 | 7.41 | 0.394 |
| 32 | Luxembourg | 0.338 | 0.383 | 9.07 | 0.424 |
| 33 | Cyprus | 0.340 | 0.380 | 8.56 | 0.415 |
| 33 | Tajikistan | 0.340 | 0.382 | 8.80 | 0.419 |
| 35 | Latvia | 0.342 | 0.396 | 10.44 | 0.444 |
| 36 | Italy | 0.354 | 0.425 | 14.28 | 0.486 |
| 37 | Portugal | 0.355 | 0.408 | 11.38 | 0.455 |
| 38 | Macedonia, FYR | 0.356 | 0.427 | 14.59 | 0.488 |
| 39 | Gambia, The | 0.359 | 0.397 | 9.57 | 0.431 |
| 39 | Romania | 0.359 | 0.428 | 14.53 | 0.488 |
| 41 | Greece | 0.360 | 0.425 | 13.79 | 0.481 |
| 41 | Thailand | 0.360 | 0.391 | 8.88 | 0.421 |
| 43 | Spain | 0.362 | 0.426 | 13.79 | 0.481 |
| 44 | Georgia | 0.365 | 0.408 | 10.69 | 0.447 |
| 45 | Lithuania | 0.374 | 0.430 | 13.62 | 0.479 |
| 46 | Tonga | 0.376 | 0.412 | 10.61 | 0.446 |
| 47 | Russian Federation | 0.377 | 0.413 | 10.61 | 0.446 |
| 48 | Myanmar | 0.381 | 0.414 | 10.57 | 0.445 |
| 49 | China | 0.386 | 0.422 | 11.31 | 0.455 |
| 50 | Ethiopia | 0.391 | 0.429 | 12.08 | 0.464 |
| 51 | Iran, Islamic Rep. | 0.395 | 0.435 | 12.75 | 0.471 |
| 52 | Indonesia | 0.397 | 0.423 | 10.80 | 0.448 |
| 53 | Philippines | 0.401 | 0.430 | 11.59 | 0.458 |
| 54 | Uruguay | 0.402 | 0.446 | 14.24 | 0.485 |
| 55 | El Salvador | 0.406 | 0.442 | 13.25 | 0.476 |
| 56 | Kenya | 0.408 | 0.443 | 13.17 | 0.475 |
| 57 | Malaysia | 0.410 | 0.446 | 13.61 | 0.479 |
| 58 | Cote d'Ivoire | 0.415 | 0.456 | 15.19 | 0.493 |
| 59 | Turkey | 0.429 | 0.466 | 15.95 | 0.500 |
| 60 | Togo | 0.431 | 0.469 | 16.63 | 0.505 |
| 61 | Peru | 0.434 | 0.484 | 20.38 | 0.529 |
| 62 | Dominican Republic | 0.452 | 0.486 | 18.37 | 0.517 |
| 63 | Ecuador | 0.460 | 0.500 | 21.81 | 0.537 |
| 64 | Bolivia | 0.467 | 0.526 | 31.64 | 0.578 |
| 65 | Paraguay | 0.476 | 0.511 | 22.94 | 0.543 |
| 66 | Chile | 0.477 | 0.510 | 22.35 | 0.540 |
| 67 | Benin | 0.478 | 0.540 | 37.60 | 0.596 |
| 68 | Costa Rica | 0.484 | 0.518 | 24.53 | 0.551 |
| 69 | Honduras | 0.496 | 0.533 | 28.46 | 0.567 |
| 70 | Panama | 0.508 | 0.551 | 35.45 | 0.590 |
| 71 | Colombia | 0.511 | 0.544 | 30.77 | 0.575 |
| 72 | Brazil | 0.513 | 0.550 | 33.67 | 0.585 |
| 73 | Botswana | 0.533 | 0.549 | 27.67 | 0.564 |
| 74 | Zambia | 0.571 | 0.592 | 44.40 | 0.613 |
| 75 | Namibia | 0.591 | 0.605 | 47.30 | 0.619 |

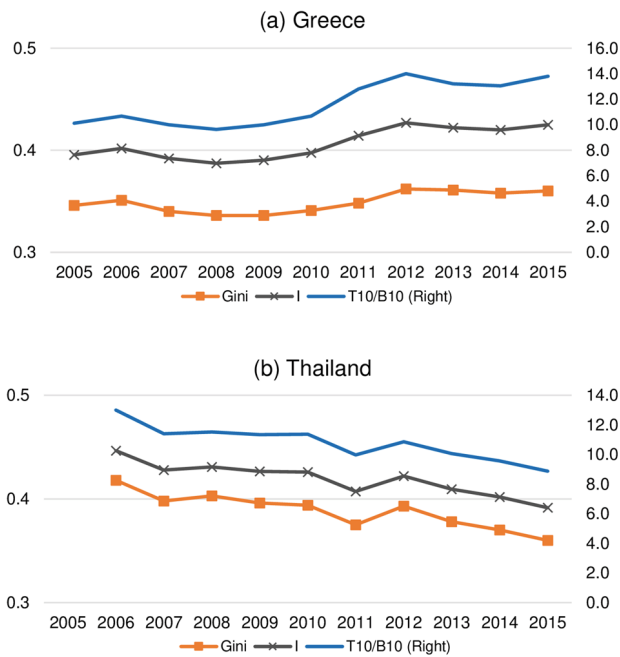

Fig. 2 The evolution of the Gini index, the $T_{10}/B_{10}$ ratio, and inequality index (I) for two countries that have the same Gini index but differ in the $T_{10}/B_{10}$ ratios (the World Bank database). a Greece. b Thailand.

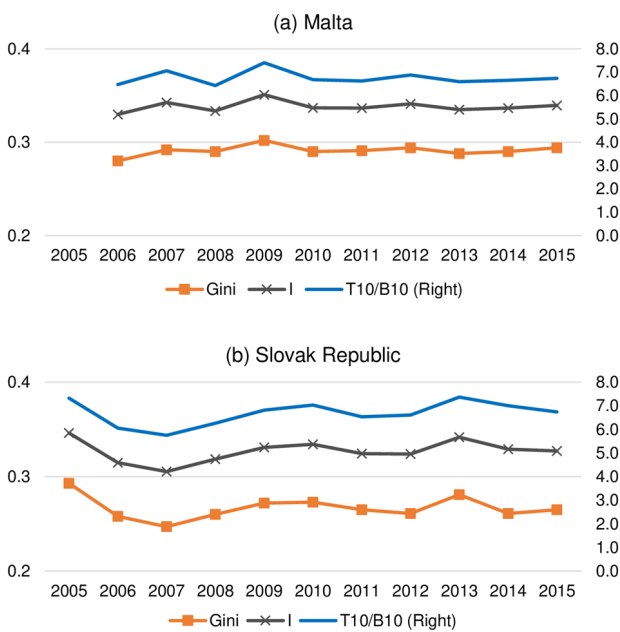

Fig. 3 The evolution of the Gini index, the $T_{10}/B_{10}$ ratio, and inequality index (I) for two countries that have the same $T_{10}/B_{10}$ ratio but differ in values of the Gini Index (the World Bank database). a Malta. b Slovak Republic.

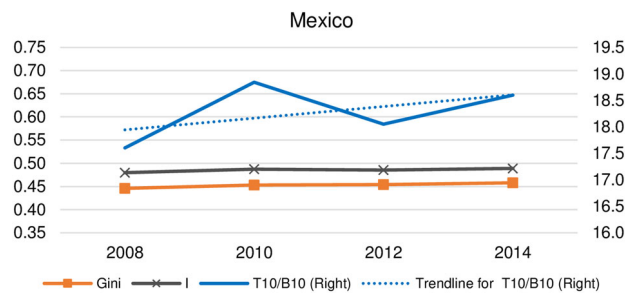

Fig. 4 The evolution of the Gini index, the $T_{10}/B_{10}$ ratio, and inequality index (I) for Mexico. It shows that the Gini index is stabilizing over time but the $T_{10}/B_{10}$ ratio is increasing (the World Bank database).

In addition to the World Bank database, we employ the data on the Gini index and the ratio of the income share of the top 10% to the income share of the bottom 10% from the OECD IDD in order to illustrate that our method still works if different database is used. Table 2 reports the ranking of countries' income inequality by the Gini index using the OECD IDD in 2015. The results confirm that our inequality index (I) would be able to distinguish the income inequality of two or more countries that share the same Gini index, but have different $\frac{T_{10}}{B_{10}}$ ratios. As discussed in the "Introduction", the United Kingdom and Israel have the same ranking of income inequality as measured by the Gini index (0.360) but the ratio of $\frac{T_{10}}{B_{10}}$ in the United Kingdom is 4.2, whereas that in Israel is 5.8, indicating that the income inequality in Israel is higher than that in the United Kingdom, which cannot be distinguished by the Gini index. Accordingly, by using our inequality index (I), it shows that the United Kingdom has I = 0.332 while Israel has I = 0.358, suggesting that income inequality in Israel is higher than that in the United Kingdom. The evolution of the Gini index, the $\frac{T_{10}}{B_{10}}$ ratio, and the inequality index (I) of the United Kingdom and Israel is illustrated in Fig. 5. In addition, our results from Table 2 show that when comparing the rankings of income inequality among countries using our inequality index (I) with those using the Gini index, there are 21 countries that their rankings have been changed while there are 14 countries whose rankings remain the same.

Using the data from the OECD IDD in 2015, our inequality index (I) could also tell the difference in income inequality when two or more countries have the same $\frac{T_{10}}{B_{10}}$ ratio, but differ in the values of the Gini index. Ireland and Switzerland share the same $\frac{T_{10}}{B_{10}}$ ratio (3.6) but the Gini index in Ireland is 0.297, whereas that in Switzerland is 0.296. This suggests that the income inequality in Ireland is slightly higher than that in Switzerland, which could be distinguished by our inequality index (I). According to our inequality index (I), Ireland has I = 0.286 while Switzerland has I = 0.285. Figure 6 shows the evolution of the Gini index, the $\frac{T_{10}}{B_{10}}$ ratio, and the inequality index (I) of Ireland and Switzerland.

Similar to the case of Mexico previously discussed where our inequality index (I) would be able to capture the dynamics of the rising of $\frac{T_{10}}{B_{10}}$ ratio over time while the Gini index is stabilizing during the same period, the data from the OECD IDD shows that Italy has somewhat stable Gini index around 0.327 between 2010 and 2014 but the $\frac{T_{10}}{B_{10}}$ ratio is rising from 4.4 in 2010 to 4.6 in 2014, indicating that the income inequality is increasing across time, which cannot be captured by the Gini index. However, our inequality index (I) could capture the dynamics of income inequality in Italy since the inequality index (I) shows a rising trend from 0.318 in 2010 to 0.322 in 2014. The evolution of the Gini index, the $\frac{T_{10}}{B_{10}}$ ratio, and the inequality index (I) of Italy is shown in Fig. 7.

Table 2 Income inequality indicators for 35 countries in 2015 (the OECD IDD).

| Ranking by Gini index | Country | Gini index | Inequality index (I) | $T_{10}/B_{10}$ | H |
|---|---|---|---|---|---|
| 1 | Slovenia | 0.250 | 0.251 | 3.2 | 0.252 |
| 2 | Slovak Republic | 0.251 | 0.249 | 3.1 | 0.246 |
| 3 | Iceland | 0.255 | 0.248 | 3.0 | 0.240 |
| 4 | Czech Republic | 0.258 | 0.252 | 3.1 | 0.246 |
| 5 | Finland | 0.260 | 0.253 | 3.1 | 0.246 |
| 6 | Denmark | 0.263 | 0.249 | 2.9 | 0.234 |
| 7 | Belgium | 0.268 | 0.266 | 3.4 | 0.264 |
| 8 | Norway | 0.272 | 0.260 | 3.1 | 0.246 |
| 9 | Austria | 0.276 | 0.267 | 3.3 | 0.258 |
| 10 | Sweden | 0.278 | 0.268 | 3.3 | 0.258 |
| 11 | Hungary | 0.284 | 0.274 | 3.4 | 0.264 |
| 12 | Netherlands | 0.288 | 0.273 | 3.3 | 0.258 |
| 13 | Poland | 0.292 | 0.292 | 4.0 | 0.293 |
| 14 | Germany | 0.293 | 0.286 | 3.7 | 0.279 |
| 15 | France | 0.295 | 0.282 | 3.5 | 0.269 |
| 16 | Switzerland | 0.296 | 0.285 | 3.6 | 0.274 |
| 17 | Ireland | 0.297 | 0.286 | 3.6 | 0.274 |
| 18 | Luxembourg | 0.306 | 0.300 | 4.0 | 0.293 |
| 19 | Canada | 0.318 | 0.314 | 4.4 | 0.310 |
| 20 | Estonia | 0.330 | 0.329 | 4.9 | 0.328 |
| 21 | Italy | 0.333 | 0.330 | 4.9 | 0.328 |
| 22 | Portugal | 0.336 | 0.329 | 4.7 | 0.321 |
| 23 | Japan | 0.339 | 0.338 | 5.2 | 0.338 |
| 24 | Greece | 0.340 | 0.336 | 5.0 | 0.331 |
| 25 | Spain | 0.345 | 0.343 | 5.3 | 0.341 |
| 26 | Latvia | 0.346 | 0.340 | 5.1 | 0.335 |
| 27 | Korea, Rep. | 0.352 | 0.352 | 5.7 | 0.353 |
| 28 | United Kingdom | 0.360 | 0.332 | 4.2 | 0.301 |
| 28 | Israel | 0.360 | 0.358 | 5.8 | 0.356 |
| 30 | Lithuania | 0.372 | 0.360 | 5.5 | 0.347 |
| 31 | United States | 0.390 | 0.377 | 6.1 | 0.364 |
| 32 | Turkey | 0.404 | 0.379 | 5.7 | 0.353 |
| 33 | Chile | 0.454 | 0.421 | 7.0 | 0.385 |
| 34 | Costa Rica | 0.479 | 0.461 | 10.3 | 0.442 |
| 35 | South Africa | 0.620 | 0.589 | 25.6 | 0.555 |

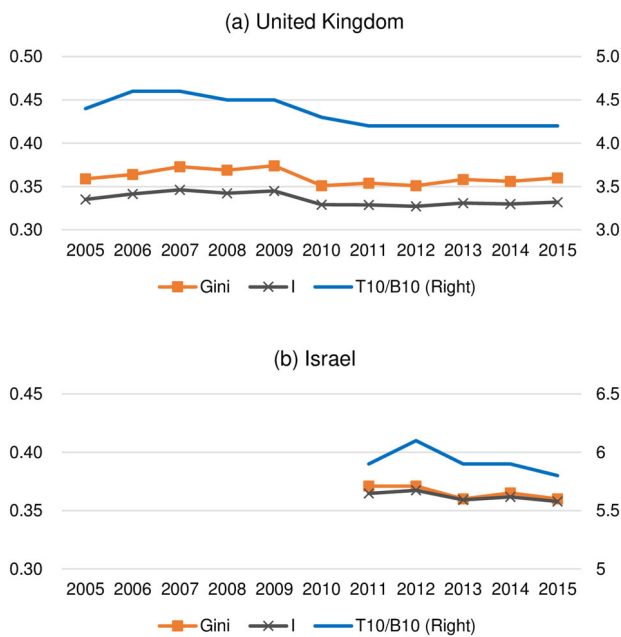

**Fig. 5 The evolution of the Gini index, the $T_{10}/B_{10}$ ratio, and inequality index (I) for two countries that have the same Gini index but differ in the $T_{10}/B_{10}$ ratios (the OECD IDD). a** United Kingdom. **b** Israel.

### Conclusions and remarks

That two or more countries have the same value of the Gini index does not necessarily imply that these countries share the same level of income inequality. In fact, income inequality could be quite different if taking into account the difference in countries' income gap between the richest and the poorest. Likewise, two or more countries having the same ratio of the income share held by the richest to the income share held by the poorest does not always imply that income inequality among these countries is the same either. The Gini index is known to be less sensitive to inequality at the tails of income distribution, whereas the ratios of income share of the richest to income share of the poorest do not account for inequality in the middle of income distribution. To overcome the limitations of the Gini index and the inter-decile ratios as measures of income inequality, this study introduces a composite index for measuring inequality that does not require the micro-data of the distribution. Our inequality index is very simple to calculate. It comprises three indicators, namely, the Gini index, the income share held by the top 10%, and the income share held by the bottom 10%. The data on these three indicators are also available, easy to access, and regularly updated by countries and international organizations.

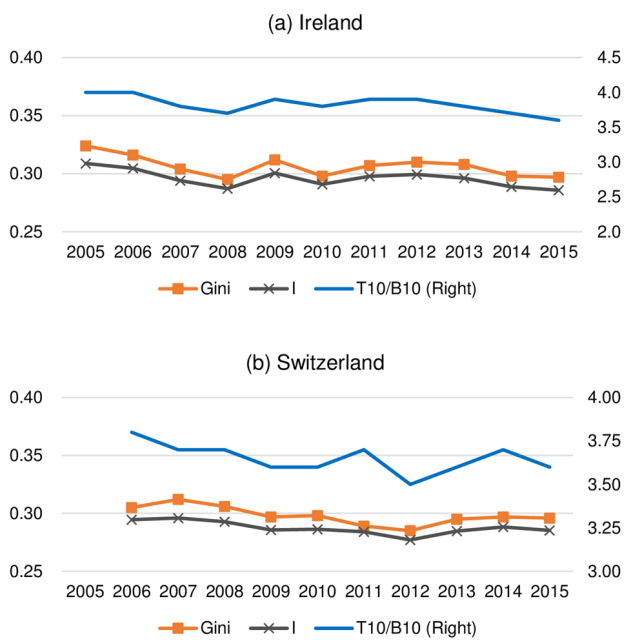

**Fig. 6 The evolution of the Gini index, the $T_{10}/B_{10}$ ratio, and inequality index (I) for two countries that have the same $T_{10}/B_{10}$ ratio but differ in values of the Gini index (the OECD IDD). a** Ireland. **b** Switzerland.

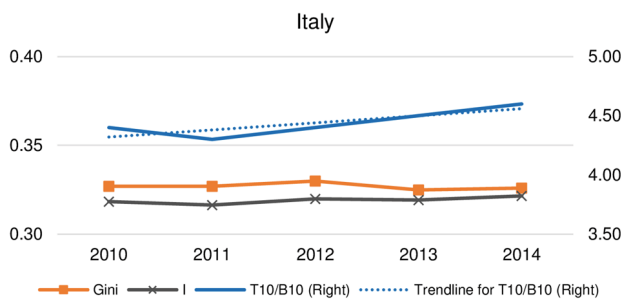

**Fig. 7 The evolution of the Gini index, the $T_{10}/B_{10}$ ratio, and inequality index (I) for Italy.** It shows that the Gini index is stabilizing over time but the $T_{10}/B_{10}$ ratio is increasing (the OECD IDD).

To demonstrate our method, we use the annual data from the World Bank and the OECD IDD between 2005 and 2015. The overall results show that our inequality index can differentiate income inequality among countries in case two or more countries share the same Gini index but differ in the income gap between the top 10% and the bottom 10%. It could also distinguish the income inequality whenever two or more countries have the same ratio of the income share of the top 10% to that of the bottom 10% but differ in values of the Gini index. In addition, our inequality index could capture the dynamics where a country's Gini index remains stable over time but the income gap between the top 10% and the bottom 10% is rising.

We would like to remark, however, that two or more countries could possibly share the same inequality index, but have different Lorenz curves as reflected by having different values of the Gini index and different ratios of the income share of the top 10% to that of the bottom 10%. Examples are Belgium and Serbia as shown in Table 1, as well as Estonia and Portugal as shown in Table 2. This implies that there are other aspects of differences in income inequality among countries that our inequality index would not be able to capture.

One way to account for other differences in income inequality is to include other inter-percentile ratios in addition to the P90/P10, say, P80/P20, P70/P30, P60/P40, and P50/P50, in the calculation of the inequality index. In this way, the whole range of the Lorenz curve would be covered. Using the same notations as before with slight modifications, for any given country i, $\left(\frac{B_x}{T_x}\right)_i$ is now defined as the ratio of the income share of the bottom $x\%$ ($B_x$) to the income share of the top $x\%$ ($T_x$), where $0 < x \leq 50$. In addition, let $j$ be the number of $x$'s, where $j = 1, 2, \ldots, N$. Next, for any given $x_j$, we have to find an appropriate value of $\alpha_{x_j}$ such that $(\text{avg. Gini}) = 1 - (\text{avg.}(\frac{B_{x_j}}{T_{x_j}}))^{\alpha_{x_j}}$ for each sample. We then use the $\alpha_{x_j}$ to calculate the $H_{ix_j}$, where $H_{ix_j} = 1 - \left(\frac{B_{x_j}}{T_{x_j}}\right)_i^{\alpha_{x_j}}$. Given the values of the $\text{Gini}_i$ and the $H_{ix_j}$ terms, the inequality index for any given country ($I_i$) can now be rewritten as a function of the $\text{Gini}_i$ and the $H_{ix_j}$ terms as follows:

$$I_i = f\left(\text{Gini}_i, H_{ix_j}\right) = \frac{\sqrt{\text{Gini}_i^2 + \sum_{j=1}^N H_{ix_j}^2}}{\sqrt{N+1}}, j = 1, 2, \ldots, N$$
$$0 \leq I_i \leq 1$$

Note that the number of $H_{ix_j}$ terms could be varied, depending upon the number of inter-percentile ratios and the availability of the data used for the calculation of the inequality index. For example, if we use five inter-percentile ratios, namely, P90/P10, P80/P20, P70/P30, P60/P40, and P50/P50 in our analysis, we would have five $H_{ix_j}$ terms, which are $H_{i10}$, $H_{i20}$, $H_{i30}$, $H_{i40}$, and $H_{i50}$. The inequality index for any given country i($I_i$) could be computed as $\frac{\sqrt{\text{Gini}_i^2 + H_{i10}^2 + H_{i20}^2 + H_{i30}^2 + H_{i40}^2 + H_{i50}^2}}{\sqrt{5+1}}$. This is one way to take into account the difference in income inequality in case two or more countries share the same inequality index but have dissimilar Lorenz curves. There might be other alternative ways to account for such a difference, which await future research.

Last but not least, we hope that our simple method for measuring inequality could be applied not only to socioeconomics, but also to broad scientific disciplines as a measure of statistical heterogeneity and for size distributions of any non-negative quantities.

### Data availability
All data generated and/or analyzed during this study are included in this manuscript and can be accessed from the World Bank and the Organization for Economic Co-operation and Development websites as listed in the references.

### Note
1 Note that an alternative and more simple index could be constructed by using only the Gini index in percentage points ($\text{Gini}_i * 100$) and the ratio of the income share held by the top 10% to that held by the bottom 10% without the need to calculate the values of the $\alpha$ and the $H_i$. The alternative index for any given country i can be defined as follows:

$$\text{Alternative index}_i = \frac{\sqrt{(\text{Gini}_i * 100)^2 + \left(\frac{T_{10}}{B_{10}}\right)_i^2}}{100}$$

This alternative index takes the values between 0.01 and ∞. When everyone has the same share of income, it is equal to 0.01 ($\text{Gini}_i = 0$ and $(T_{10}/B_{10})_i = 1$). When one person has all incomes and everyone else has none, the alternative index is equal to ∞ ($\text{Gini}_i = 1$ and $(T_{10}/B_{10})_i = \infty$ since $B_{10} = 0$), making it difficult to interpret and compare among countries as discussed in the "Introduction".

### Acknowledgements
T.S. is grateful to Dr. Suradit Holasut for guidance and comments.

### Competing interests
The authors declare no competing interests.

### Additional information
**Supplementary information** is available for this paper at https://doi.org/10.1057/s41599-020-0484-6.

**Correspondence** and requests for materials should be addressed to T.S.

**Reprints and permission information** is available at http://www.nature.com/reprints

**Publisher's note** Springer Nature remains neutral with regard to jurisdictional claims in published maps and institutional affiliations.